\documentclass[preprint,showpacs,showkeys,aps,prb]{revtex4}
\usepackage[dvips]{graphicx}

\begin{document}

\title{
Ergodicity of a Time-Reversibly Thermostated \\
Harmonic Oscillator and the 2014 Ian Snook Prize
}

\author{
William Graham Hoover and Carol Griswold Hoover               \\
Ruby Valley Research Institute                  \\
Highway Contract 60, Box 601                    \\
Ruby Valley, Nevada 89833                       \\
}

\date{\today}


\vspace{0.1cm}

\begin{abstract}
Shuichi Nos\'e opened up a new world of atomistic simulation in 1984.  He formulated
a Hamiltonian tailored to generate  Gibbs' canonical distribution dynamically.
This clever idea bridged the gap between microcanonical molecular dynamics and
canonical statistical mechanics. Until then the canonical distribution was
explored with Monte Carlo sampling. Nos\'e's dynamical Hamiltonian bridge
requires the ``ergodic'' support of a space-filling structure in order to reproduce
the entire distribution.  For sufficiently small systems, such as the harmonic
oscillator, Nos\'e's dynamical approach failed to agree with Gibbs' sampling
and instead showed a complex structure, partitioned into a chaotic sea, islands, and
chains of islands, that is familiar textbook fare from investigations of Hamiltonian chaos.
In trying to enhance small-system ergodicity several more complicated
``thermostated" equations of motion were developed.  All were consistent with the
canonical Gaussian distribution for the oscillator coordinate and momentum.  The
ergodicity of the various approaches has undergone several investigations, with
somewhat inconclusive ( contradictory ) results. Here we illustrate {\it several} ways
to test ergodicity and challenge the reader to find even more convincing algorithms or
an entirely new approach to this problem.

\end{abstract}

\maketitle

\section{Introduction}
In 1989 Shuichi Nos\'e won IBM-Japan's science prize for his 1984 discovery\cite{b1,b2} of
a dynamical Hamiltonian approach to Gibbs' canonical ensemble.  He added a new ``time-scaling''
variable $s$ ( and its conjugate momentum $p_s$ ) to the usual list of $\#$ Cartesian
degrees of freedom, coupling the kinetic energy to a logarithmic temperature-dependent
potential for $s$ :
$$
{\cal H}_N(q,p,s,p_s) = \sum^\# (p^2/2ms^2) + \Phi(q) + (p_s^2/2M) + (\# +1)kT\ln(s) \ .
$$
Multiplying the resulting equations of motion by $s$ ( ``scaling the time'' )
gives a set of $(2\# + 2)$ dynamical motion equations fully consistent with Gibbs'
canonical ensemble ,
$$
f(q,p) \propto e^{-{\cal H}/kT} \ ,
$$
provided that we replace $(p/s)$ by $p$ in the scaled equations of motion.

Hoover\cite{b3} soon pointed out the lack of ergodicity for this approach, applied to a harmonic
oscillator,  and at the same time simplified the derivation of a set of $(2\# + 1)$
equations of motion only slightly different to Nos\'e's :
$$
\{ \ \dot q = (p/m) \ ; \ \dot p = F - \zeta p \ \} \ ;
 \ \dot \zeta = \sum^\# [ \ (p^2/mkT) - 1 \ ]/\tau^2 \ .
$$
 The friction coefficient $\zeta$ in these ``Nos\'e-Hoover'' equations is proportional
to Nos\'e's $p_s$ . These equations are {\it much} better behaved numerically and also
make the extra $s$ variable redundant. Carl Dettmann showed\cite{b4} that these same equations
follow from a Hamiltonian like Nos\'e's provided that the Dettmann Hamiltonian is
arbitrarily set equal to zero and the number of degrees of freedom is $\#$ rather than
$\# + 1$ ,
$$
{\cal H}_D \equiv s{\cal H}_N \equiv 0 \ .
$$

Hoover and Harald Posch and Franz Vesely\cite{b5,b6} investigated the details of these dynamics for the
harmonic oscillator and discovered an infinite variety of periodic, toroidal, and chaotic
solutions, with the sum of all of these disparate parts equal to the canonical distribution,
the product of three Gaussian functions, in $q$, in $p$, and in $\zeta$ for the oscillator.

Bauer, Bulgac, and Kusnezov\cite{b7,b8} generalized the thermostating approach to include two or more
control variables, even managing to reproduce Brownian motion by using three of them.
Soon after, Martyna, Klein, and Tuckerman\cite{b9} suggested the use of a chain of thermostat
variables to enhance ergodicity.  Their equations of motion ( for the shortest chain and
with all the free parameters equal to unity ) are :
$$
\{ \ \dot q = p \ ; \ \dot p = -q - \zeta p \ ;
 \ \dot \zeta = p^2 - 1 - \xi \zeta \ ; \ \dot \xi = \zeta^2 - 1 \ \} \ . \ [ \ {\rm MKT} \ ]
$$
Presently, Hoover and Holian suggested a different dual-control approach\cite{b10}, fixing both the
second and fourth velocity moments :
$$
\{ \ \dot q = p \ ; \ \dot p = -q - \zeta p - \xi p^3\ ; \
\dot \zeta = p^2 - 1 \ ; \ \dot \xi = p^4 - 3p^2 \ \} \ . \ [ \ {\rm HH} \ ]
$$
Equations controlling the sixth or higher moments tend to be too stiff for practical use.

Very recently Patra and Bhattacharya\cite{b11} took a different direction, stimulated by work on the
``configurational temperature'', as opposed to the usual kinetic one.  For the oscillator 
the configurational temperature\cite{b12} is proportional to the potential energy, $\langle \ q^2 \ \rangle$.
Patra and Bhattacharya had the clever idea of imposing both temperatures simultaneously on
the oscillator.  Their feedback equations use one friction coefficient, $\zeta$ , to control
kinetic temperature, and the other, $\xi$ , to control configurational temperature ,
$\langle \ q^2 \ \rangle $ :
$$
\{ \ \dot q = p - \xi q\ ; \ \dot p = -q - \zeta p \ ; \
\dot \zeta = p^2 - 1 \ ; \ \dot \xi = q^2 - 1 \ \} \ . \ [ \ {\rm PB} \ ]
$$
We wish to emphasize that all three approaches followed here, Martyna-Klein-Tuckerman,
Hoover-Holian, and Patra-Bhattacharya, if ergodic, provide {\it exactly the same}
four-dimensional Gaussian distribution ,
$$
(2\pi)^2f(q,p,\zeta,\xi) \equiv e^{-(q^2/2)}e^{-(p^2/2)}e^{-(\zeta^2/2)}e^{-(\xi^2/2)} \ .
$$
To prove that this is so it is only necessary to evaluate Liouville's continuity equation for
the flow in phase space, showing that the four-dimensional Gaussian distribution is stationary :
$$
(\partial f/\partial t) \equiv -f[ \ (\partial \dot q/\partial q) +(\partial \dot p/\partial p)
+ (\partial \dot \zeta/\partial \zeta) + (\partial \dot \xi/\partial \xi) \ ]
$$
$$
- \dot q(\partial f/\partial q) - \dot p(\partial f/\partial p) -
\dot \zeta(\partial f/\partial \zeta) -
\dot \xi(\partial f/\partial \xi) \equiv 0 \ .
$$

Let us turn to the question of establishing ergodicity for the three dynamical models, MKT, HH,
and PB .  Our own approach, because we know it best, is computational.  It is possible that a
more convincing ``rigor mortis'' approach could be developed from a mathematical standpoint.  At
the moment {\it any} ``proof'' of ergodicity is an ``open problem''.  In seven brief sections we
describe some of the methods that have been applied to this problem\cite{b13}, ending with a
challenge for the reader.

\begin{figure}[h]
\includegraphics[width=8.0cm,angle=+90]{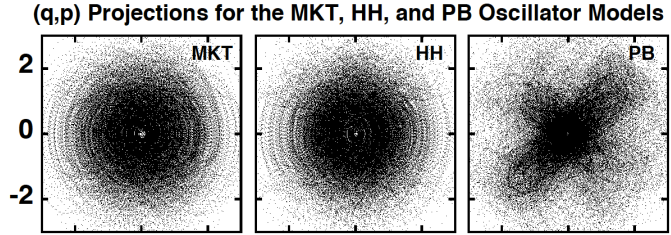}
\vspace{1cm}
\caption{
Chaotic sea for the Martyna-Klein-Tuckerman, Hoover-Holian, and
Patra-Bhattacharya oscillators.  The initial condition for these trajectories was
$(q,p,\zeta,\xi) = (0,5,0,0)$ in all three cases.  100,000 equally spaced $(q,p)$ points are
plotted for each model.  The Patra-Bhattacharya oscillator
has clear deviations from hyperspherical symmetry.  The fourth-order Runge-Kutta timestep
is 0.001 for all of the computations in our work.  Here ten million timesteps were used.
Note the holes in the MKT and HH projections and the lack of circular symmetry in the PB
projection.
}
\label{Figure1}
\end{figure}

\section{Time averages of the moments}
Because kinetic temperature is the second velocity moment it is usual to confirm that the first few
velocity moments agree with the canonical distribution.  A short fourth-order Runge-Kutta calculation,
adding $(p^2, p^4, p^6)$ to the list of righthandsides being integrated, shows that the second, fourth,
and sixth moments of the MKT and HH equations match the values expected from a Gaussian,
(1, 3, 15) to three or four significant figures while the PB model gives instead
(1.000, 3.8, 23.5) .  These results are conclusive evidence that only the MKT and HH equations
are candidates for ergodicity. Nevertheless we will apply each of our numerical tests to all
three sets of motion equations, in order to see the difference between the difficult and the easy.

\section{Projections}
The Projections of the four-dimensional dynamics onto {\it any} plane, such as the
$(q,p),(q,\zeta),(q,\xi),(p,\zeta),(p,\xi),(\zeta,\xi)$ planes, should look Gaussian
{\it without any holes}.  For
simplicity we use the initial condition $p=5$ with the other variables initially zero, promoting
chaos for all three sets of equations. The MKT and HH equations both appear to have a small hole
near the origin of the $(qp)$ plane.  The PB equations produce a very different pattern, showing
a strong positive correlation between the two variables, another indication that these equations
though sometimes chaotic, are not ergodic.  See {\bf Figure 1} for a comparison of the three models in the
$(q,p)$ planar projection.  These results suggest two further checks, a look at the density
near the $(q,p)$ origin and a search for periodic orbits, which would be necessary in order that any
holes form in the Gaussian distributions.

\section{Density}
The probability density for the four-dimensional Gaussian at a radius of $0.1$ has decreased
from its maximum value at the $(0,0,0,0)$ origin by a factor of  
$$
[ \ \rho(r=0.1)/\rho(r=0.0) \ ] = e^{-r^2/2} = e^{-0.005} \simeq 1.00000 - 0.00500 + 0.00001 = 0.99501 \ ,
$$
so that the densities determined from trajectories should hardly vary within
that four-dimensional sphere {\it provided that the flow is ergodic}.  Measuring density requires
considerably longer runs because the time interval between visits near the origin increases at
least as rapidly as $(r^{-4}) $ .  We measured the probability densities for the MKT and HH
algorithms inside $(q,p)$ circles of area $0.01\pi \times(1,1/2,1/4,1/8,1/16)$ using a billion timesteps, and
found no significant difference in density over that range.  Evidently the MKT and HH algorithms
behave as though they were ergodic.  The difference between those densities and the PB chaotic
density was approximately a factor of two, indicating that approximately half the PB measure is
chaotic and the other half quasiperiodic, and so omitted from this chaotic simulation.

It is worth mentioning that the MKT equations have two fixed points, though in the end neither one
is actually attractive.  In the special case that $q$ and $p$ both vanish so that $\dot q$ and
$\dot p$ are likewise zero, the remaining equations of motion are :
$$
\dot \zeta = -1 - \xi \zeta \ ; \ \dot \xi = \zeta ^2 - 1 \ .
$$
These equations describe a flow from the unstable two-dimensional fixed point $(-1,+1)$ to the
stable ( only in two dimensions ) fixed point $(+1,-1)$ .  These flow equations are isomorphic
to those of a one-dimensional particle in a constant field, with $\zeta$ playing the r\^ole of
momentum and $\xi$ acting as the friction coefficient.  Patra and Bhattacharya's assertion that
the MKT equations are not ergodic relies on analyses in the vicinity of these two fixed points.
That type of work is complicated by the fact that the four-dimensional flow becomes more and more
intermittent as one nears either of these two points.  Nevertheless we believe that our analysis
in this Section and particularly in Section VII below casts doubt on their claim ( hence our
offering of the 2014 Ian Snook Prize on that subject, as is described in Section IX ) .

\section{Periodic Orbits}
Choosing an initial condition somewhat closer to the origin
$$
\{ \ q,p,\zeta,\xi \ \} = \{ \ 0.68,0.68,0,0 \ \}
$$ 
gives $(q,p,\zeta )$ projection plots which ``look'' spherical for the MKT and HH equations.  The PB
equations, on the other hand, provide a clearly-defined torus, shown in {\bf Figure 2} . The PB dynamics
has been projected into three-dimensional $(q,p,\zeta)$ space.  The PB projection shows that the
phase-space distribution has at least one large cavity in the chaotic sea and that the cavity contains
a family of nested tori.

\vspace{1cm}
\begin{figure}[ht]
\includegraphics[width=5.cm,angle= 90]{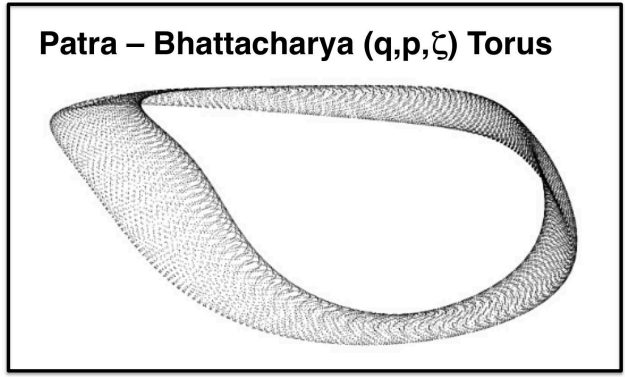}
\caption{
Patra-Bhattacharya torus with initial conditions $(q,p,\zeta,\xi) = (0.68,0.68,0,0)$.
Choosing (0.5,0.5,0.0,0.0) generates a very thin torus close to the periodic orbit at the center of
the larger torus shown in the figure. 50,000 points are plotted.
}
\label{Figure2}
\end{figure}

\section{Ensemble Tests}
Because a lack of ergodicity implies segregated regions in the $(q,p,\zeta,\xi)$ space it is a
tempting idea to study ensembles of initial conditions, expecting to find two or more distinct
longtime-average values rather than a Gaussian Central-Limit-Theorem distribution around a single
ergodic average value. In addition to moments, and their correlations, the largest Lyapunov exponent
should be a particularly ``good'' property to follow.  Although some complicated attractors ( such as
Rayleigh-B\'enard flows ) have more than a single chaotic strange-attractor region such a situation
is implausible for the simple oscillator.  Therefore it seems likely that a random or a grid-based
selection of initial points would produce at least a bimodal distribution of values for a nonergodic
set of equations.  Some ensemble tests of this kind were carried out in References 6 and 13.

\section{Many Lyapunov Exponents}
\begin{figure}[ht]
\includegraphics[angle= 90,scale=.4]{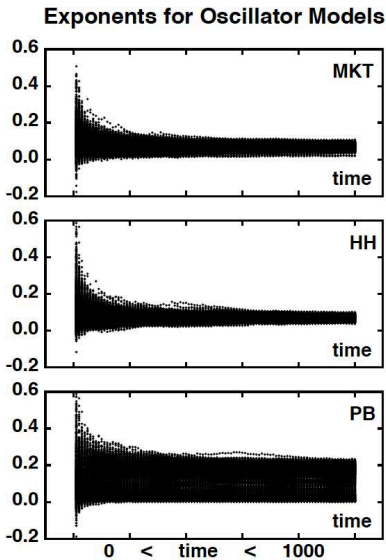}
\caption{
Time-averaged Lyapunov exponents for the Martyna-Klein-Tuckerman, Hoover-Holian, and
Patra-Bhattacharya oscillators. 1000 trajectories are traced from points located randomly within a
$ 4 \times 4 \times 4 \times 4 $ hypercube centered at the coordinate origin.  Although there is no
reason that the MKT and HH results should agree, or nearly so, the two models have similar Lyapunov
exponents.  The Patra-Bhattacharya motion equations lead to two clumps of exponents at long times,
one at 0.0000 and the other at 0.14.
}
\label{Figure3}
\end{figure}

Taking up the ensemble idea we choose 1000 different initial conditions in the four-dimensional
hypercube of sidelength four centered on the origin, following each of them for $10^6$ timesteps.  Though
there is no reason for the HH and MKT exponents to agree there turns out to be good agreement between
them, as {\bf Figure 3} shows.  On the other hand the PB equations' Lyapunov exponents separate into two
distinct values, 0.0000 and 0.14, characteristic of their separate portions of the phase space.  We view the density
and Lyapunov tests as the most convincing evidence that both the MKT and the HH algorithms {\it are}
ergodic.

\section{History of the Extremal Lyapunov Exponents}
\begin{figure}[ht]
\includegraphics[angle=90,scale=.3]{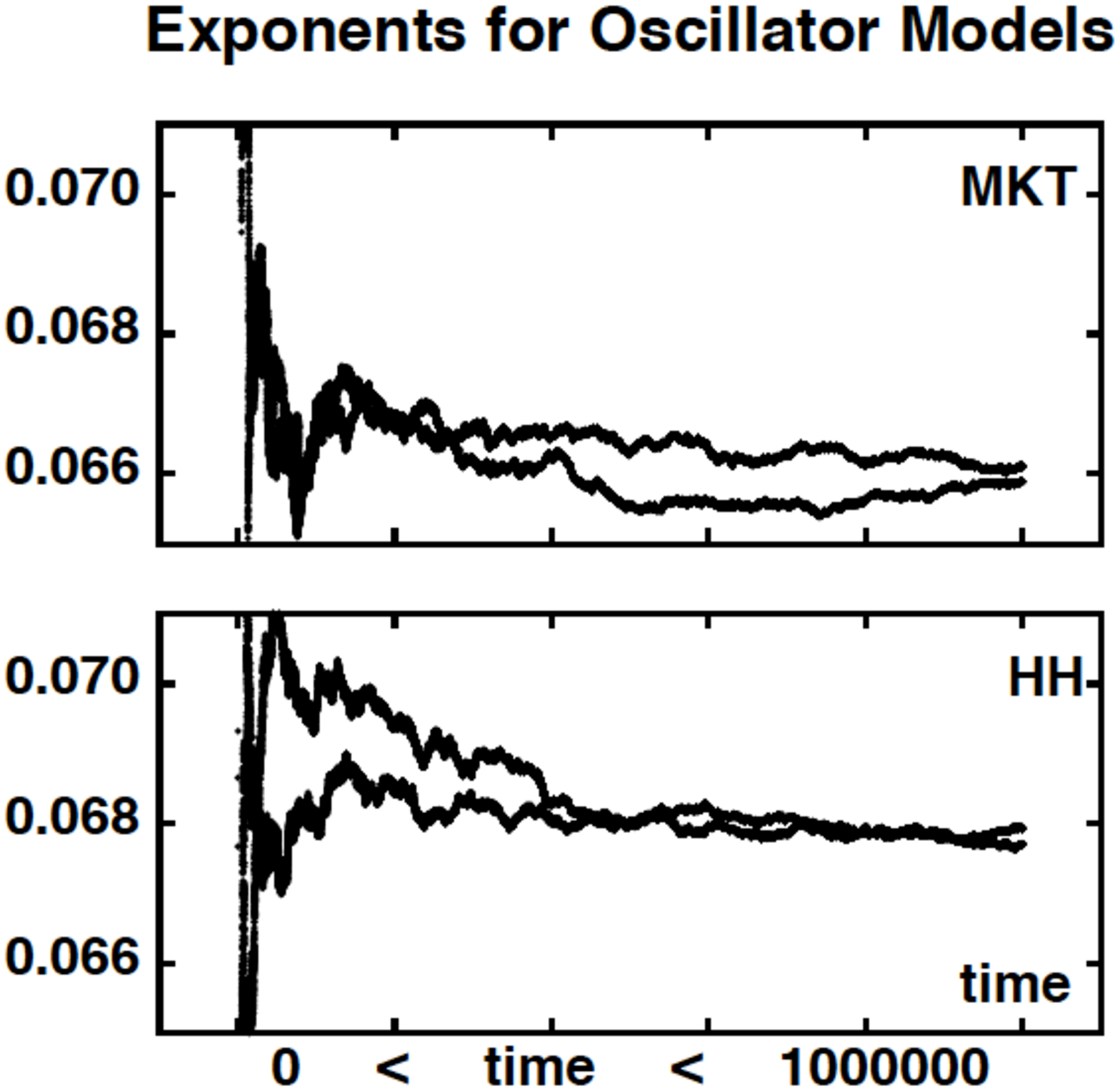}
\caption{
Time-averaged Lyapunov exponents for two runs 1000 times longer than those in Figure 3.  The initial
conditions were chosen to match those of the minimum and maximum values shown at the righthand
margin of Figure 3. The convergence of the two outliers toward a common value is a strong indicator of
ergodicity.
}
\label{Figure4}
\end{figure}

To nail this conclusion down we took the two extreme cases ( the maximum and minimum Lyapunov
exponent out of 1000 simulations ) and ran them 1000 times longer.  {\bf Figure 3} shows the behavior
of the ensemble of initial conditions for the Martyna-Klein-Tuckerman and Hoover-Holian models.
{\bf Figure 4} shows a comparison of the maximum and minimum Lyapunov-exponent calculations, carried out
for a time 1000 times longer.  The MKT and HH data show no significant difference between them,
strongly suggesting that the entire set of random trajectories started in a $ 4 \times 4 \times 4 \times 4 $
hypercube are part of the same chaotic sea, filling all of space in a Gaussian manner for the 
Martyna-Klein-Tuckerman and Hoover-Holian oscillators.

\section{Further Tests}
Patra and Bhattacharya also determined phase-space density numerically, in one- or two-dimensional
cross sections\cite{b15}.  They also computed mean deviations of
long-time-averaged distributions from the canonical one as averages.  Their final conclusion
is that the Martyna-Klein-Tuckerman algorithm is not ergodic.  Our
own evidence for that same problem points in the opposite direction, toward ergodicity.  We
thank Puneet Patra, Baidurya Bhattacharya, and Clint Sprott for an exchange of hundreds of
stimulating emails on this general subject, and specially appreciated a personal visit from
Professor Bhattacharya this summer.

\section{Conclusions -- Ian Snook Prize for 2014}
The disagreement between our own investigations past and present ( which agree with those of
Martyna, Klein, and Tuckerman ) and those of Patra and Bhattacharya are both thought-provoking
and stimulating. Last year we offered a prize to honor the memory of our Australian Colleague
Ian Snook. We asked arXiv readers for a time-reversed version of a simple pseudorandom number
generator with a cycle length of $2^{22}$ .  Within 24 hours Professor Federico Ricci-Tersenghi
( University of Rome ) won the prize with the elegant solution described in his arXiv.1305.1805
contribution\cite{b16}, ``The Solution to the Challenge in `Time-Reversible Random Number Generators' by
Wm. G. Hoover and Carol G. Hoover''.  His generator can be used, for instance, to reverse
Brownian dynamics trajectories, a demonstration problem in our forthcoming book on the
management and control of nonequilibrium systems.

\vspace{2.5cm}
\begin{figure}[ht]
\includegraphics[width=12.0cm,angle= 90]{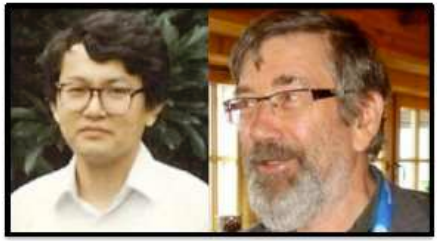}
\caption{
Shuichi Nos\'e ( 1951-2005 ) and Ian Snook ( 1945-2013 )
}
\label{Figure5}
\end{figure}

This year we invite the readers of Computational Methods in Science and Technology to consider
thoughtfully this interesting problem in classical statistical mechanics and dynamical systems
theory.  Specifically the Ian Snook Prize for 2014 will be awarded to whomever submits the 
most convincing treatment of this problem to us prior to 1 January 2015.  The question to
be discussed is ``To what extent are trajectory-based solutions of the Martyna-Klein-Tuckerman
harmonic-oscillator motion equations ergodic?''

It is our intention to reward the most convincing entry received ( or submitted to the arXiv
or to Computational Methods in Science and Technology )  prior to 1 January 2015.  The 2014
Ian Snook prize of five hundred United States dollars will be presented to the winner in
January 2015.  We would be very grateful for your contributions.  We dedicate this work to
the memories of our two colleagues shown in {\bf Figure 5} .

\section{Acknowledgments}
We thank Puneet Patra and Baidurya Bhattacharya for bringing this problem
to our attention and additionally thank Clint Sprott for his insights into a wide
variety of fascinating problems in dynamical systems theory and for his thoughts
on the current prize offer.


\begin{thebibliography}{99}
\bibitem{b1}  S. Nos\'e, ``A Unified Formulation of the Constant Temperature Molecular Dynamics Methods'',
              Journal of Chemical Physics {\bf 81}, 511-519 (1984).
\bibitem{b2}  S. Nos\'e, ``Constant Temperature Molecular Dynamics Methods'', Progress in Theoretical
              Physics Supplement {\bf 103}, 1-46 (1991).
\bibitem{b3}  Wm. G. Hoover, ``Canonical Dynamics: Equilibrium Phase-Space Distributions'', Physical
              Review A {\bf 31}, 1695-1697 (1985).
\bibitem{b4}  C. P. Dettmann and G. P. Morriss, ``Hamiltonian Formulation of the Gaussian Isokinetic
              Thermostat'', Physical Review E {\bf 54}, 2495-2500 (1996).
\bibitem{b5}  H. A. Posch, W. G. Hoover, and F. J. Vesely, ``Canonical Dynamics of the Nos\'e Oscillator:
              Stability, Order, and Chaos'', Physical Review A {\bf 33}, 4253-4265 (1986).
\bibitem{b6}  H. A. Posch and Wm. G. Hoover, ``Time-Reversible Dissipative Attractors in Three and
              Four Phase-Space Dimensions'', Physical Review E {\bf 55}, 6803-6810 (1997).
\bibitem{b7}  D. Kusnezov, A. Bulgac, and W. Bauer, ``Canonical Ensembles from Chaos'', Annals of Physics
              {\bf 204}, 155-185 (1990).
\bibitem{b8}  D. Kusnezov and A. Bulgac, ``Canonical Ensembles from Chaos: Constrained Dynamical
              Systems'', Annals of Physics {\bf 214}, 180-218 (1992).
\bibitem{b9}  G. J. Martyna, M. L. Klein, and M. Tuckerman, ``Nos\'e-Hoover Chains---the Canonical Ensemble
              {\it via} Continuous Dynamics'', Journal of Chemical Physics {\bf 97}, 2635-2643 (1992).
\bibitem{b10} Wm. G. Hoover and B. L. Holian, ``Kinetic Moments Method for the Canonical Ensemble
              Distribution'', Physics Letters A {\bf 211}, 253-257 (1996).
\bibitem{b11} P. K. Patra and B. Bhattacharya, ``A Deterministic Thermostat for Controlling
              Temperature using All Degrees of Freedom'', Journal of Chemical Physics {\bf 140}, 064106 (2014).
\bibitem{b12} K. P. Travis and C. Braga, ``Configurational Temperature and Pressure Molecular Dynamics:
              Review of Current Methodology and Applications to the Shear Flow of a Simple Fluid'',
              Molecular Physics {\bf 104}, 3735-3749 (2006).
\bibitem{b13} Wm. G. Hoover, Carol Hoover, and Dennis Isbister, ``Chaos, Ergodic Convergence, and Fractal
              Instability for a Thermostated Canonical Harmonic Oscillator'',  Physical Review E {\bf 63},
              3541-3546 (2000).
\bibitem{b14} Wm. G. Hoover and C. G. Hoover, ``Time-Reversible Random Number Generators :
              Solution of Our Challenge by Federico Ricci-Tersenghi'': arXiv.1305.0961.
\bibitem{b15} Puneet Kumar Patra and Baidurya Bhattacharya, ``Non-Ergodicity of Nos\'e-Hoover Chain Thermostat
              in Computationally Achievable Time'': arXiv.1407.2353.
\bibitem{b16} F. Ricci-Tersenghi, ``The Solution to the Challenge in `Time-Reversible Random Number
              Generators' by Wm. G. Hoover and Carol G. Hoover'': arXiv.1305.1805.

\end{thebibliography}
\end{document}